\documentclass[3p, preprint, review,times]{elsarticle}

\usepackage{lineno,hyperref}
\usepackage{url}

\usepackage{textcomp}
\usepackage{xcolor}
\usepackage{amsmath,amssymb,amsfonts}
\usepackage{algorithmic}
\usepackage{graphicx}
\usepackage{subcaption}

\modulolinenumbers[5]

\biboptions{sort&compress}

\begin{document}

\begin{frontmatter}

\title{Adaptive Sign Algorithm for 
Graph Signal Processing}
\author[1]{Yi Yan
}
\ead{y-yan20@mails.tsinghua.edu.cn}
\author[1]{Ercan E. Kuruoglu\corref{cor1}
}
\ead{kuruoglu@sz.tsinghua.edu.cn}
\author[2]{Mustafa A. Altinkaya
}
\ead{mustafaaltinkaya@iyte.edu.tr}

\address[1]{Tsinghua Berkeley Shenzhen Institute, Tsinghua University, Shenzhen, China.}
\address[2]{Izmir Institute of Technology, Izmir, Turkey}
\cortext[cor1]{Corresponding author}

\begin{abstract}
Efficient and robust online processing technique of irregularly structured data is crucial in the current era  of data abundance. In this paper, we propose a graph/network version of the classical adaptive Sign algorithm for online graph signal estimation under impulsive noise. Recently introduced graph adaptive least mean squares algorithm is unstable under non-Gaussian impulsive noise and  has high computational complexity. The Graph-Sign algorithm proposed in this work is based on the minimum dispersion criterion and therefore impulsive noise does not hinder its estimation quality. Unlike the recently proposed graph adaptive least mean p-th power algorithm, our  Graph-Sign algorithm can operate without prior knowledge of the noise distribution. The proposed Graph-Sign algorithm has a faster run time because of its low computational complexity compared to the existing adaptive graph signal processing algorithms. Experimenting on steady-state and time-varying graph signals estimation utilizing spectral properties of bandlimitedness and sampling, the Graph-Sign algorithm demonstrates fast,  stable, and robust graph signal estimation performance under impulsive noise modeled by alpha stable, Cauchy, Student’s t, or Laplace distributions.
\end{abstract}


\end{frontmatter}

\section{Introduction}
Graph-based data structures are gaining popularity in recent years due to the effective  power of graphs in representing multivariate irregular data in fields such as data science, information engineering, bioinformatics, and finance \cite{b1,b2,b3,bib_graphML}. 
However, with this increasing popularity of the utilization of graphs, the traditional data processing techniques that were optimized on structured data could not adapt to the structural irregularities and could not utilize the intrinsic relationships among data seen in or modelled by graphs, which led to a demand for algorithms that could  process graph-structured data efficiently \cite{b1,b2,b3}. 
The recently emerged Graph signal processing (GSP) techniques are efficient solutions to deal with the previously mentioned irregularities in real applications such as in modeling brain structure \cite{brain_modeling}, monitoring 5G Signal Strength \cite{bib_LMS}, modelling nationwide temperature \cite{bib_NLMS}, monitoring sensor networks in smart cities \cite{smart_city}, structuring geometric data \cite{bib_geometric}, and modeling transportation flows \cite{b_7_trafic}. 
GSP techniques are also seen in the foundation of spectral graphical deep learning algorithms, such as ChebNet \cite{bib_cheb} and graph convolutional networks \cite{bib_GCN}, where nonlinear activation functions are applied to the data processed through GSP techniques to incorporate non-linearity. The GSP components in the graph neural-networks provide model interpretability, which was previously lacking in the non-graphical deep learning algorithms \cite{bib_graphML}.
GSP-based algorithms have the capability of solving many classical machine learning tasks such as classification \cite{bib_GCN} and clustering \cite{bib_clustering}. 
However, these algorithms can only handle static tasks that do not operate in real-time; this drawback of the static models urges the need for online data processing techniques.

In classical signal processing, online estimation of time-varying signals is often accomplished using adaptive filters \cite{bib_classical}. Adaptive GSP algorithms are inspired by classical adaptive filters to perform online estimation of steady-state and time-varying graph signals through spectral methods \cite{bib_LMS,bib_NLMS,b10,bib_LMP}. 
Analogous to the famous adaptive least mean squares (LMS) algorithm in classical adaptive filtering, the GSP least mean squares algorithm (GLMS) is popular due to its simplicity of modeling the noise using Gaussian distribution and using $l_2$-norm optimization to estimate the output \cite{bib_LMS}. 
Other least-squares based adaptive GSP algorithms such as the GSP normalized LMS (GNLMS) \cite{bib_NLMS} and the GSP recursive least squares (GRLS) algorithm \cite{bib_RLS} are extensions from classical adaptive filtering algorithms and have been introduced to reduce the number of iterations till convergence. 
To tackle the time-varying nature of some real-world data, another line of work known as Time-vertex Signal Processing was proposed to take into account the time-domain information. The cost function of time-vertex signal processing algorithms is set up using similar to the GLMS $l_2$-norm cost function with an additional regularization term and utilizes Joint Time-Vertex Fourier Transform for the inclusion of the time domain information \cite{bib_time_vertex}.

The ambient noise in various real-world applications is non-Gaussian with impulsive characteristics that can be modeled by heavy-tailed distributions such as Student's t, generalized Gaussian, and $\alpha$-stable distributions \cite{bib_nonGaussian}. 
For example, in underwater communications the noise is a Cauchy-Gaussian mixture \cite{bib_underwater}, in powerline communications the noise is modeled using $\alpha$-stable \cite{b17_PLC}. 
However, least-squares based algorithms, namely the GLMS, the GRLS, and the GNLMS algorithms, assume the noise to be Gaussian noise, which is invalid for impulsive noise situations. The least-squares optimization is unstable and will diverge under the presence of large or infinite variance caused by the impulsiveness in the noise \cite{b18_lp_ref}. 
The GRLS algorithm further requires prior knowledge of the covariance matrix, which is not acquirable in many cases. 
In order to avoid the drawbacks caused by $l_2$-norm optimization when the noise distribution is impulsive and non-Gaussian, the GSP least mean $p^{th}$ power algorithm (GLMP) further assumes the noise to be symmetric $\alpha$-stable distribution (S$\alpha$S). Under the S$\alpha$S noise assumption, the optimal cost function of GLMP is derived based on the minimum dispersion criterion, leading to $l_p$-norm optimization instead of $l_2$-norm optimization \cite{bib_LMP}. 
However, there are two drawbacks of the GLMP algorithm: it has additional computations to the already expensive GLMS algorithm, and its parameter selection is still based on the prior knowledge of the noise.

In this paper, we propose the adaptive Graph-Sign algorithm (G-Sign) as a graph extension of the classical adaptive sign error or the least mean absolute deviation (LMAD) algorithm for multivariate signals \cite{bib_classical,bib_MD_LMAD}. The proposed G-Sign algorithm is derived based on the minimum dispersion criterion and then reduced to $l_1$-norm optimization, which removes the need for prior knowledge from any noise assumption \cite{bib_l1}. This allows the G-Sign algorithm to avoid the instability seen in the least-squares-based algorithms when estimating graph signals under impulsive noise. Compared to the GLMP and the GLMS algorithms the G-Sign algorithm further reduces computational complexity to estimate a steady-state graph signal,
making the G-Sign algorithm time efficient. 
The G-Sign algorithm is robust when estimating time-varying graph signal under impulsive noise, making it capable to perform online graph signal estimation. 
Note that the GSP based adaptive algorithms update the graph-signal estimates instead of the weights for the data samples as done by classical adaptive algorithms. 
The desired parallelism of the inference in predicting the outcomes of typically multiple hundreds of nodes is made possible by the graph version of the signal processing algorithms. 
Some dimensionality reduction is achieved by localizing the graph signal both spatially and spectrally. 
However, introducing large amount of weights which needs to be jointly optimized is mainly avoided by preferring data-centered approach in graph-based gradient methods.

The remaining sections of this paper are organized as follows. The background information on GSP and noise modeling are in Section \ref{sec:Background}. In \ref{sec:Alrorithm}, we derive the G-Sign algorithm and analyze its computational complexity. The experimental studies and results are in \ref{sec:results}. Section 5 provides the conclusions.

\section{Background}
\label{sec:Background}
\subsection{GSP Preliminaries}
A graph $\mathcal{G}=(\mathcal{V}, \mathcal{E})$  is defined with
with a set of  $N$ nodes $\mathcal{V} = v_1 ... v_N$, and a set of $M$  edge $\mathcal{E}=  e_1 ... e_M $ representing the connections between nodes. In this paper, we assume that the graph $\mathcal{G}$ is undirected and can be either weighted or unweighted. 
A graph signal $\boldsymbol{x}$ is a graph with a function value defined on the nodes. The adjacency matrix $\mathbf{A}$ of the graph $\mathcal{G}$ represents the connectivity of the edges in $\mathcal{E}$. The $ij^{th}$ entry of $\mathbf{A}$ is the edge weight from node $v_i$ to node $v_j$ when $\mathcal{G}$ is weighted or simply $\mathbf{A_{ij} = 1}$ when there is an edge between node $v_i$ and node $v_j$ when $\mathcal{G}$ is unweighted. For an undirected graph, the adjacency matrix $\mathbf{A}$ is symmetric.
If $\mathcal{G}$ is undirected and unweighted, the number of edges a node $v_i$ has is the node degree $d_i$, where we can formulate a diagonal matrix called the degree matrix $\mathbf{D} = $ diag$(d_1...d_N)$. 
In the weighted case, the degree of a node is the summation of all of the edge weights instead of simply counting the number of edges.
The graph Laplacian matrix $\mathbf{L}$ combines the information from $\mathbf{A}$ and $\mathbf{D}$ is defined simply as $\mathbf{L=D-A}$. 

The graph Fourier transform (GFT) is defined based on $\mathbf{L}$ by performing the eigenvector decomposition $\mathbf{L=U\Lambda U^\mathit{T}}$, where $\mathbf{U}$ is a matrix composed of the orthonormal eigenvectors of $\mathbf{L}$, and $\mathbf{\Lambda}$ is a diagonal matrix of eigenvalues $\boldsymbol{\lambda} = [\lambda_1, ... ,\lambda_N]^\mathit{T}$. 
The GFT transforms the graph signal $\boldsymbol{x}$ from  spatial-domain to spectral domain by projecting $\boldsymbol{x}$ onto $\mathbf{U}$:  $\boldsymbol{s}=\mathbf{U}^\mathit{T}\boldsymbol{x}$.
Spectral-domain operations could be performed similarly as in the classical Fourier Transform case by defining a filter $\mathbf{H(\boldsymbol{\lambda})}$ and then applied using the convolution property of Fourier Transform.
A graph signal transformed to the spectral domain could utilize the inverse graph Fourier transform (IGFT) $\mathbf{\boldsymbol{x}=U\boldsymbol{s}}$ to transform back to the spatial domain.
Here is a basic yet complete GSP procedure to apply the filter to $\boldsymbol{x}$ and generate a processed graph signal $\boldsymbol{x}_p = \mathbf{UH}(\boldsymbol{\lambda})\mathbf{U}^\mathit{T}\boldsymbol{x}.$
Define a frequency set $\mathcal{F}$, a bandlimiting filter $\mathbf{\Sigma}$ has $\mathbf{\Sigma_{ii}} = 1$ if the $i^{th}$ frequency is to be included in $\mathcal{F}$ and 0 otherwise, then the filter $H(\boldsymbol{\lambda})$ is $    \mathbf{H}(\boldsymbol{\lambda}) = \mathbf{\Sigma} =  \text{diag}\left(\boldsymbol{1}_\mathcal{F}\left(\boldsymbol{\lambda}\right)\right)$, where $\boldsymbol{1}_\mathcal{F}\left(\lambda_i\right) = 1$ if $\lambda_i\in\mathcal{F}$ and $0$ otherwise. 
A graph signal is sparse in spectral sense when it is bandlimited in the spectral domain. 
A graph signal with reduced number of nodes sampled based on a sampling node set $\mathcal{S}\subseteq\mathcal{V}$ is sparse in the spatial domain \cite{b10}; $\mathbf{D}_\mathcal{S}$ is the sampling matrix with $\mathbf{D}_\mathcal{S\text{ii}} = 1$ $\forall \, v_i \in \mathcal{S}$.
$\mathbf{D}_\mathcal{S}$ and $\Sigma$ are idempotent and self-adjoint matrices.

\subsection{Impulsive Distributions}
To reflect real-life noises which are sometimes impulsive and non-Gaussian, we use the following non-Gaussian distributions to model the noise: the S$\alpha$S, the Cauchy, the Student's t, and the Laplace \cite{bib_nonGaussian}. 
The S$\alpha$S is a distribution that satisfies the generalized central limit theorem and governed by the characteristic exponent $\alpha$, the location parameter $\mu$, and the scale parameter $\gamma$.  The mean of S$\alpha$S is only defined when $\alpha>1$: with $\mu$ being the mean when $1<\alpha\leqslant2$, and the median when $\alpha<1$. The variance of S$\alpha$S is defined only when $\alpha$ = 2, and in other cases the concept of dispersion is used instead \cite{bib_MD_LMAD}. Gaussian distribution is obtained by setting  $\alpha = 2$, and Cauchy distribution when $\alpha = 1$; the S$\alpha$S has no analytic PDF for other $\alpha$ values, but has the characteristic function \begin{equation}
    \boldsymbol{\phi}(t)=\exp\left\{j\mu t-\gamma|t|^\alpha\right\}.\label{SaS}
\end{equation}

The Cauchy distribution is a special case of the S$\alpha$S at $\alpha = 1$, and a special case of Student's t distribution with the $\nu=1$. 
The Cauchy distribution is heavy-tailed with variance undefined and the PDF is
\begin{equation}
    \boldsymbol{f}(t,\gamma)=\frac{1}{\pi\gamma\left[1+\left(\frac{t-\mu_\alpha}{\gamma}\right)^2\right]}.\label{cauchy}
\end{equation}

The Student's t distribution is governed by its degrees of freedom $\nu$, with  infinite variance when $1<\nu\leqslant2$, and undefined variance when $\nu<1$. The Student's t distribution becomes the Gaussian distribution when $\nu = \infty$. The PDF of Student's t distribution is
\begin{equation}
    \boldsymbol{f}(t)=\frac{\Gamma\left(\frac{\nu+1}{2}\right)}{\sqrt{\nu\pi}\Gamma\left(\frac{\nu}{2}\right)}\left(1+\frac{t^2}{\nu}\right)^{\frac{-\nu+1}{2}}, \label{student_t}
\end{equation}
where $\Gamma$ is the gamma function. 

Finally, the Laplace distribution is a special case of generalized Gaussian distribution governed by the location parameter $\mu$ and the scale parameter $b$, with PDF
\begin{equation}
    \boldsymbol{f}(t,\mu,b)=\frac{1}{2b}\exp\left(-\frac{|t-\mu|}{b}\right).\label{laplace}
\end{equation}

\section{The Adaptive G-Sign Algorithm}
\label{sec:Alrorithm}
\subsection{Algorithm Derivation and Complexity Analysis}
\label{sec:Alrorithm_1}

Let's consider a bandlimited graph signal $\boldsymbol{x_0}\subseteq\mathbb{R}^N$ and its noisy observation with missing nodes $\boldsymbol{y}\left[k\right] = \mathbf{D_\mathcal{S}}\left(\boldsymbol{x_0}+\boldsymbol{w}\left[k\right]\right)$, where partial observations are modeled using a sampling matrix $\mathbf{D_\mathcal{S}}$, and $k$ represents the $k^{th}$ time step or iteration. 
$\boldsymbol{w}\left[k\right]$ is a zero-mean noise and it is i.i.d. among the nodes and across the time.
Computation $\mathbf{U\Sigma U}^\mathit{T}$ can be reduced by defining  $\mathbf{U_\mathcal{F}=U\Sigma}$ then droping the all zeros columns, resulting in $\mathbf{U_\mathcal{F}U_\mathcal{F}^\mathit{T}\boldsymbol{x} = B\boldsymbol{x}}$ \cite{b10}. For a perfectly bandlimited graph signal with frequency bands $\mathcal{F}$, $\mathbf{\boldsymbol{x} = B\mathbf{\boldsymbol{x}}}$ \cite{b10}. 
In GLMS, the current step estimate of $\boldsymbol{x_0}$ is $\hat{\boldsymbol{x}}\left[k\right]$; $\hat{\boldsymbol{x}}\left[k\right]$ could be obtained by solving a convex optimization problem in which the cost function could be formed to minimize the error between $\boldsymbol{y}\left[k\right]$ and $\hat{\boldsymbol{x}}\left[k\right]$:
\begin{equation}
        J\left(\hat{\boldsymbol{x}}\left[k\right]\right)=\mathbb{E}\left\Vert\boldsymbol{y}\left[k\right]-\mathbf{D_\mathcal{S}B\hat{\boldsymbol{x}}\left[k\right]}\right\Vert_2^2. \label{lms_cost} 
\end{equation}
In order to make one-step ahead prediction, the  spatial-domain update could be derived by stochastic gradient descent:
\begin{equation}
    \hat{\boldsymbol{x}}\left[k+1\right]=\hat{\boldsymbol{x}}\left[k\right]+\mu_{lms} \mathbf{BD_\mathcal{S}(\boldsymbol{y}\left[k\right]-\hat{\boldsymbol{x}}\left[k\right])},
    \label{lms_update}
\end{equation}
where $\mu_{lms}$ is the step-size. 

Even though the GLMS algorithm is simple, it is not optimal in time efficiency and estimation stability. The GLMP algorithm is an extension of the GLMS algorithm that has stable estimation performance compared to GLMS when estimating a graph signal under S$\alpha$S noise but with additional complexity\cite{bib_LMP}. 
In classical adaptive filtering, the LMS algorithm is used extensively due to its simplicity of implementation, and the Sign-Error algorithm or the LMAD algorithm is an extension of the LMS algorithm to further increase run-speed and to decrease algorithm complexity, with additional robustness gained from the $l_1$-norm cost function. To improve time-efficiency and robustness of adaptive GSP algorithms under impulsive noise, we use the minimum dispersion criterion to form the cost function and reduce it to a $l_1$-norm optimization problem similar to the approaches from the LMAD algorithm in classical adaptive filtering \cite{bib_MD_LMAD}:
\begin{equation}
        J\left(\hat{\boldsymbol{x}}\left[k\right]\right)=\mathbb{E}\left\Vert\boldsymbol{y}\left[k\right]-\mathbf{D_\mathcal{S}B\hat{\boldsymbol{x}}\left[k\right]}\right\Vert_1^1. \label{sign_cost} 
\end{equation}
When a distribution is S$\alpha$S, Cauchy, Laplace, or Student's t, such $l_1$-norm optimization is the optimal choice for parameter estimation when the density parameters are unknown \cite{bib_nonGaussian}.
The cost function \eqref{sign_cost} can be viewed as recovering the mean $\boldsymbol{x_0}$ from distribution $\boldsymbol{y}[k]$ and is LMAD sense optimal for for S$\alpha$S and Cauchy noise. Equation \eqref{sign_cost} is also the optimal Maximum Likelihood Estimator for parameter estimation in Laplace distribution.
Using the bandlimitedness property $\mathbf{B}\hat{\boldsymbol{x}}\left[k\right] = \hat{\boldsymbol{x}}\left[k\right]$, the update function of the G-Sign algorithm obtained by stochastic gradient as in \eqref{lms_update}
\begin{equation}
    \begin{split}
    \hat{\boldsymbol{x}}\left[k+1\right]&=\hat{\boldsymbol{x}}\left[k\right]-\mu_{s}\frac{\partial \left\Vert\boldsymbol{y}\left[k\right]-\mathbf{D_\mathcal{S}B\hat{\boldsymbol{x}}\left[k\right]}\right\Vert_1^1}{\partial \hat{\boldsymbol{x}}\left[k\right]}
    \\
    &=\hat{\boldsymbol{x}}\left[k\right]+\mu_{s} \mathbf{BD}_\mathcal{S}
    \text{Sign}\left(\mathbf{D}_\mathcal{S}\left(\boldsymbol{y}\left[k\right]-\hat{\boldsymbol{x}}\left[k\right]\right)\right).
    \label{sign_update_1}
    \end{split}
\end{equation}
The Sign(.) function in the update equation results from taking the derivative of the $l_1$ norm cost function with the consideration of the point of discontinuity of the derivative at 0. This makes the update function resemble  the form seen in the weight update of classical LMAD or Sign-Error algorithm \cite{bib_MD_LMAD,bib_classical}. A step size parameter $\mu_{s}$ is added by following classical adaptive filtering convention to control the amount of update. The $i^{th}$ 0 in the diagonal of $\mathbf{D}_\mathcal{S}$ corresponds to a 0 in the $i^{th}$ element of sign$(\mathbf{D}_\mathcal{S}\left(\boldsymbol{y}\left[k\right]-\hat{\boldsymbol{x}}\left[k\right])\right)$. So, we can safely refactor \eqref{sign_update_1} into 
\begin{equation}
    \hat{\boldsymbol{x}}\left[k+1\right]
    =\hat{\boldsymbol{x}}\left[k\right]+\mu_{s} \mathbf{B}
    \text{Sign}\left(\mathbf{D}_\mathcal{S}\left(\boldsymbol{y}\left[k\right]-\hat{\boldsymbol{x}}\left[k\right]\right)\right).
    \label{sign_update_2}
\end{equation}

This fixed amount of update from the minimum dispersion criterion is unaffected by impulsive noise \cite{bib_MD_LMAD}. In the GRLS algorithm the update contains the covariance matrix of the noise \cite{bib_RLS}. In the GLMP algorithm the exponent $p$ of the update term is determined based on $\alpha$ of the S$\alpha$S noise \cite{bib_LMP}. 
Unlike the algorithms that select the parameters using prior information from noise statistics, the G-Sign algorithm requires no prior information to determine the only parameter $\mu_{s}$, which is in correspondence with the classical LMAD or the Sign-Error algorithm \cite{bib_l1}.

Equation \eqref{sign_update_2} reduces the number of operations by $2(N-|\mathcal{S}|)$ 
so the zeros in $\mathbf{D}_\mathcal{S}$ make $\text{Sign}\left(\mathbf{D}_\mathcal{S}\left(\boldsymbol{y}\left[k\right]-\hat{\boldsymbol{x}}\left[k\right]\right)\right)$ sparse. 
The Sign($\cdot$) operation essentially compares the non-zero elements in $\mathcal{S}$, where in the worst case $\boldsymbol{y}\left[k\right] = \hat{\boldsymbol{x}}\left[k\right]$ it compares all the digits of $\hat{\boldsymbol{x}}\left[k\right]$ and $\boldsymbol{y}\left[k\right]$, but does no mathematical computation. Because $\hat{\boldsymbol{x}}\left[k\right]$ is a noisy estimation of an observation $\boldsymbol{y}\left[k\right]$, this worst case $\hat{\boldsymbol{x}}\left[k\right] = \boldsymbol{y}\left[k\right]$ is unlikely to happen. 
The analysis of the computational complexity of our G-Sign algorithm compared to the GLMS and the GLMP algorithm is in Table \ref{complexity}.
\begin{table}[htbp]
    \caption{Computational Complexity Analysis}
    \centering
    \begin{tabular}{cccc}
    \hline
         & GLMS & GLMP & G-Sign\\
    \hline
        Addition & $N^2+N$ & $N^2+2N$ & $N^2+|S|$ \\
        Multiplication & $N^2+2N$ & $N^2+3N$ & $N^2+N+|S|$ \\
        $p^{th}$ power &  0 & $N$ & 0 \\
        Sign($\cdot$) & 0 & $N$ & $|S|$\\
    \hline
    \end{tabular}
        \label{complexity}
    
\end{table}
\subsection{Mean-Squared Stability Analysis Under Steady State Estimation}
To estimate the steady-state performance of G-Sign algorithm, the mean-squared deviation (MSD) at step $k$ is being calculated: 
\begin{equation}
    \text{MSD}\left[{k}\right]=\frac{1}{N} {\left\Vert\hat{\boldsymbol{x}}\left[k\right]-\boldsymbol{x_0}\right\Vert}_2^2.
\end{equation} 
Let the error of estimating $\boldsymbol{x_0}$ at step $k$ be $\tilde{\boldsymbol{x}}[k] = \boldsymbol{\hat{x}}[k]-\boldsymbol{x_0}$, then the error of the update  \eqref{sign_update_2} is
\begin{equation}
    \tilde{\boldsymbol{x}}\left[k+1\right]
    =\tilde{\boldsymbol{x}}\left[k\right]+\mu_{s}\mathbf{B}
    \text{Sign}\left(\mathbf{D}_\mathcal{S}\left(\boldsymbol{w}\left[k\right]-\mathbf{U_\mathcal{F}}\tilde{\boldsymbol{s}}\left[k\right]\right)\right).
    \label{sign_error_1}
\end{equation}
Using GFT to transform \eqref{sign_error_1} in to the spectral domain \eqref{sign_error_1} becomes 
\begin{equation}
    \tilde{\boldsymbol{s}}\left[k+1\right]
    =\tilde{\boldsymbol{s}}\left[k\right]+\mu_{s}\mathbf{U}_\mathcal{F}^T\mathbf{D}_\mathcal{S}\mathbf{R}\left(\boldsymbol{w}\left[k\right]-\mathbf{U}_\mathcal{F}\tilde{\boldsymbol{s}}\left[k\right]\right),
    \label{sign_error_2}
\end{equation}
with $\mathbf{R}\approx\text{diag}\left(|\boldsymbol{w}\left[k\right]|^{.-1}\right)$ considering that $\boldsymbol{w}\left[k\right]\gg\mathbf{U}_\mathcal{F}\tilde{\boldsymbol{s}}\left[k\right]$ when $k$ is large, and $(.)^{.-1}$ is the element wise inverse. 
The error or the deviation from the ground-truth value \eqref{sign_error_2} in mean-squared sense is
\begin{equation}
     \mathbb{E}\Vert\tilde{\boldsymbol{s}}\left[k+1\right]\Vert^2 =\mathbb{E}\Vert\tilde{\boldsymbol{s}}\left[k\right]\Vert^2_\mathbf{\Phi}+\mu_s^2\mathbb{E}\Vert\mathbf{U}_\mathcal{F}^T\mathbf{D}_\mathcal{S}\mathbf{R}\boldsymbol{w}\left[k\right]\Vert^2.
\label{MSD_error}   
\end{equation}
where $\mathbf{\Phi} = \left(\mathbf{I}-\mu_s\mathbf{U}^T_\mathcal{F}\mathbf{D}_\mathcal{S}\mathbf{RU}_\mathcal{F}\right)^T\left(\mathbf{I}-\mu_s\mathbf{U}^T_\mathcal{F}\mathbf{D}_\mathcal{S}\mathbf{RU}_\mathcal{F}\right)$, and $\Vert\tilde{\boldsymbol{s}}\left[k\right]\Vert^2_\mathbf{\Phi}$ is the weighted Euclidean norm $\tilde{\boldsymbol{s}}^T\left[k\right]\mathbf{\Phi}\tilde{\boldsymbol{s}}\left[k\right]$. Equation\eqref{MSD_error} can be factorized using the Trace-Trick $\mathbb{E}\{\mathbf{X^TYX}\} = $Tr $(\mathbb{E}\{\mathbf{XX^TY}\})$:
\begin{equation}
    \mathbb{E}\Vert\tilde{\boldsymbol{s}}\left[k+1\right]\Vert^2 =\mathbb{E}\Vert\tilde{\boldsymbol{s}}\left[k\right]\Vert^2_\mathbf{\Phi}+\mu_s^2\text{Tr}\left(\mathbf{U}_\mathcal{F}^T\mathbf{D}_\mathcal{S}\mathbf{CD}_\mathcal{S}\mathbf{U}_\mathcal{F}\right),
        \label{MSD_error_2}  
\end{equation}
where $\mathbf{C} = \mathbb{E}\Vert\mathbf{R}\boldsymbol{w}\left[k\right]\Vert^2$ is the covariance matrix of $\mathbf{R}\boldsymbol{w}\left[k\right]$ and has a simillar structure to the partial correlation matrix.
Because we assumed in \ref{sec:Alrorithm_1} that the noise among each node is i.i.d., $\mathbf{C} = \mathbf{I}$, combining with the idempotent and self-adjoint property of the sampling matrix $\mathbf{D}_\mathcal{S}$, \eqref{MSD_error_2} can be simplified to 
\begin{equation}
     \mathbb{E}\Vert\tilde{\boldsymbol{s}}\left[k+1\right]\Vert^2=\mathbb{E}\Vert\tilde{\boldsymbol{s}}\left[k\right]\Vert^2_\mathbf{\Phi}+\mu_s^2\text{Tr}\left(\mathbf{U}_\mathcal{F}^T\mathbf{D}_\mathcal{S}\mathbf{U}_\mathcal{F}\right).
     \label{MSD_error_3}
\end{equation}

Utilizing the property Tr$\{\mathbf{YX}\}=$vec$\left(\mathbf{X}^T\right)$vec$\left(\mathbf{Y}\right)$, \eqref{MSD_error_3} can be expressed in this recursive relationship:
\begin{equation}
    \mathbb{E}\Vert\tilde{\boldsymbol{s}}\left[k+1\right]\Vert^2 = \mathbb{E}\Vert\tilde{\boldsymbol{s}}\left[0\right]\Vert^2_\mathbf{\Phi^k}+\mu^2_s\text{vec}\left(\mathbf{G}\right)^T\sum_{i=0}^{k}\mathbf{\Phi}^i,
    \label{MSD_error_4}  
\end{equation}
where $\mathbf{G=U}^T_\mathcal{F}\mathbf{D_\mathcal{S}U_\mathcal{F}}$ and $\tilde{\boldsymbol{s}}\left[0\right]$ is the error at step $k=0$. From \eqref{MSD_error_4}, we see that $\mathbb{E}\Vert\tilde{\boldsymbol{s}}\left[k+1\right]\Vert^2$ converges to a steady value if the RHS of \eqref{MSD_error_4} converges. 
When the initial error $\tilde{\boldsymbol{s}}\left[0\right]$ is bounded, the RHS of \eqref{MSD_error_4} is a constant value when summation term becomes a geometric series, leading to the condition $\Vert(\mathbf{I}-\mu_{s}\mathbf{U}^T_\mathcal{F}\mathbf{D}_\mathcal{S}\mathbf{RU_\mathcal{F}})\Vert < 1$ for steady-state convergence of the algorithm. 
For a diagonalizable matrix $\mathbf{X}$ of size $N \times N$and a vector $\mathbf{z}$ of size $N$, $\Vert\mathbf{Xz}\Vert^2 = \sum \lambda_i z_i$ which could result in $\Vert\mathbf{X}\Vert \leq |\lambda_{max}|$. where $\lambda_{i = 1...N}$ are the eigenvalues of $\mathbf{X}$ and $\lambda_{max}$ is the largest eigenvalue of $\mathbf{X}$. 
By letting $\mathbf{X} = \mathbf{U}^T_\mathcal{F}\mathbf{D}_\mathcal{S}\mathbf{RU_\mathcal{F}}$, if $0 < |1-\mu_{s}\lambda_{max}| < 1$, then $\Vert(\mathbf{I}-\mu_{s}\mathbf{U}^T_\mathcal{F}\mathbf{D}_\mathcal{S}\mathbf{RU_\mathcal{F}})\Vert < 1$ is satisfied.
$\mathbf{R}$ in $\mathbf{\Phi}$ could be approximated using then the fractional lower order moment of $|\boldsymbol{w}\left[k\right]|^{-1}$, resulting in $\mathbf{R} {\approx} \mathbb{E}|\boldsymbol{w}\left[k\right]|^{-p_s}\mathbf{I}$, with $p_s = 0.99$ being a number that is slightly smaller than 1 \cite{bib_MD_LMAD}. 
As a result, $\mu_s$ should satisfy 
\begin{equation}
    0 < \mu_s < \frac{2}{\lambda_\text{max}},
    \label{bound}
\end{equation}
where $\lambda_\text{max}$ is the maximum eigenvalue of $\mathbf{U_\mathcal{F}}^T\mathbf{D}_\mathcal{S}\mathbf{R}\mathbf{U_\mathcal{F}}$. 
The condition \eqref{bound} has a structure similar to the condition seen in GLMS, but the the bound $2/\lambda$ is obtained using a different expression than GLMS; this bound is derived based on the Sign($\cdot$) part of the update function \eqref{sign_update_2}.
Under this condition, the G-Sign algorithm is stable and the MSD behavior will converge to a bounded value for a steady state estimation.

As $k\to\infty$ and $\mu_s$ satisfies \eqref{bound}, $\lim_{k \to \infty} \tilde{\boldsymbol{s}}\left[k\right] = \lim_{k \to \infty} \tilde{\boldsymbol{s}}\left[k+1\right]$.
The theoretical steady-state MSD can be calculated by rewriting \eqref{MSD_error_3} using the property vec$\left(\mathbf{XYZ}\right) = \left(\mathbf{Z}^T \otimes \mathbf{X}\right)$vec$\left(\mathbf{Y}\right)$ then use the energy conservation approach of GFT \cite{bib_LMS} to find:
\begin{equation}
    \begin{split}
        \text{MSD}\left[k\right] & =\lim_{k \to \infty} \mathbb{E}\Vert\tilde{\boldsymbol{x}}\left[k\right]\Vert^2 = \lim_{k \to \infty} \mathbb{E}\Vert\tilde{\boldsymbol{s}}\left[k\right]\Vert^2 \\
        & =\mu_{s}^2\text{vec}\left(\mathbf{G}\right)^T(\mathbf{I-Q})^{^-1}\text{vec}\left(\mathbf{I}\right),
    \end{split}  
    \label{MSD_simulation}
\end{equation}
where $Q = \left(\mathbf{I}-\mu_s\mathbf{U}^T_\mathcal{F}\mathbf{D}_\mathcal{S}\mathbf{RU}_\mathcal{F}\right)^T\otimes\left(\mathbf{I}-\mu_s\mathbf{U}^T_\mathcal{F}\mathbf{D}_\mathcal{S}\mathbf{RU}_\mathcal{F}\right)$.
\subsection{Absolute Deviation Stability Analysis Under Steady State Estimation}
Instead of the MSD analysis  in section \ref{sec:steady_state}, we further derived the requirement for the G-Sign algorithm to converge for a steady-state estimation using the Mean Absolute Deviation (MAD) at step $k$:
\begin{equation}
    \text{MAD}\left[{k}\right]=\frac{1}{N} {\left|\hat{\boldsymbol{x}}\left[k\right]-\boldsymbol{x_0}\right|}.
\end{equation} 

We first start from the spectral-domain update error derived in \eqref{sign_error_2}; it can be factored into the following error update equation:
\begin{equation}
    \begin{split}
    \tilde{\boldsymbol{s}}\left[k+1\right] & = \left(\mathbf{I}-\mu_s\mathbf{U}^T_\mathcal{F}\mathbf{D}_\mathcal{S}\mathbf{R}\mathbf{U}_\mathcal{F}\right)\tilde{\boldsymbol{s}}\left[k\right]+\mu_s\mathbf{U}^T_\mathcal{F}\mathbf{D}_\mathcal{S}\mathbf{R}\boldsymbol{w}[k-1]\\
    & = \mathbf{\Phi}^k_{1}\tilde{\boldsymbol{s}}\left[0\right]+\mu_s\sum_{i=0}^{k}\mathbf{\Phi}^{k}_{1}\mathbf{U}^T_\mathcal{F}\mathbf{D}_\mathcal{S}\mathbf{R}\boldsymbol{w}[k-i],\label{MAD}
    \end{split}
\end{equation}
where $\mathbf{\Phi}_1 = \left(\mathbf{I}-\mu_s\mathbf{U}^T_\mathcal{F}\mathbf{D}_\mathcal{S}\mathbf{R}\mathbf{U}_\mathcal{F}\right)$.
By taking the limit of the expected absolute value of \eqref{MAD} as $k\to\infty$ and using the same approximation for $\mathbf{R}$ in section \ref{sec:steady_state}, the spectral domain MAD update can be expressed as
\begin{equation}
    \lim_{k \to \infty} \mathbb{E}\left|\tilde{\boldsymbol{s}}\left[k\right]\right| = \lim_{k \to \infty} \mathbb{E}\left|\mathbf{\Phi}^k_1\tilde{\boldsymbol{s}}\left[0\right]+\mu\sum_{i=0}^{k}\mathbf{\Phi}^i_1\mathbf{U}^T_\mathcal{F}\mathbf{D}_\mathcal{S}\boldsymbol{1}\right|,
    \label{MAD2}  
\end{equation}
where $\boldsymbol{1}$ is an all-ones vector of size $N\times1$. 
To let the RHS of \eqref{MAD2} converge, it requires $\Vert(\mathbf{I}-\mu_{s}\mathbf{U}^T_\mathcal{F}\mathbf{D}_\mathcal{S}\mathbf{RU_\mathcal{F}})\Vert < 1$, so that the summation becomes a geometric series and $\mathbf{\Phi}^k_1\tilde{\boldsymbol{s}}\left[0\right]$ approximates to $0$ for a bounded $\tilde{\boldsymbol{s}}\left[0\right]$. This condition is the same condition as the MSD case in section \ref{sec:steady_state}, which will lead to \eqref{bound}. 
\section{Experimental Results}
\label{sec:results}
\begin{figure}[htbp]
\centerline{\includegraphics[width=0.5\textwidth]{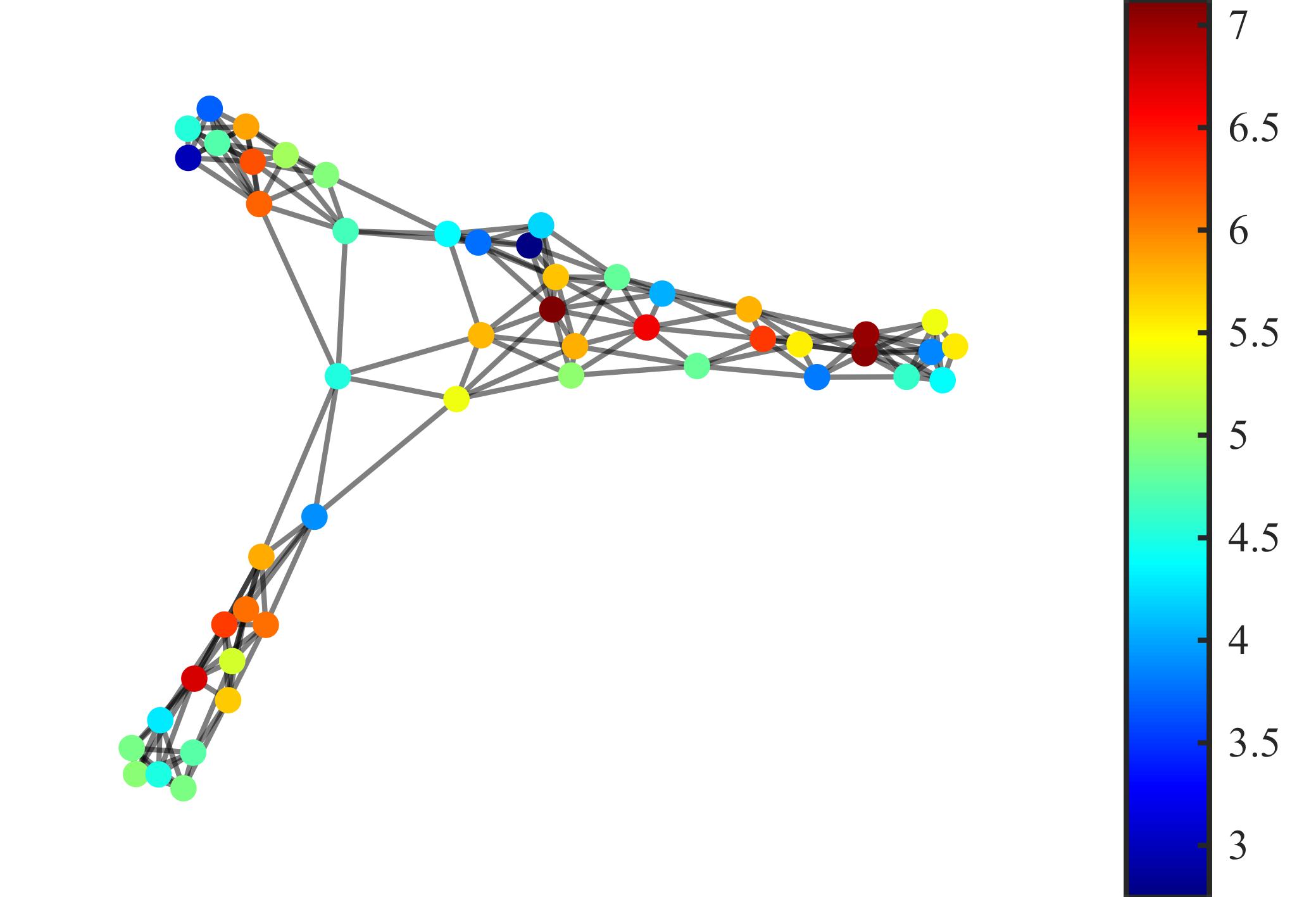}}
\caption{The graph signal of a sensor graph and its topology.}
\label{top_1}
\end{figure}

\begin{figure}[htbp]
\centerline{\includegraphics[width=0.5\textwidth]{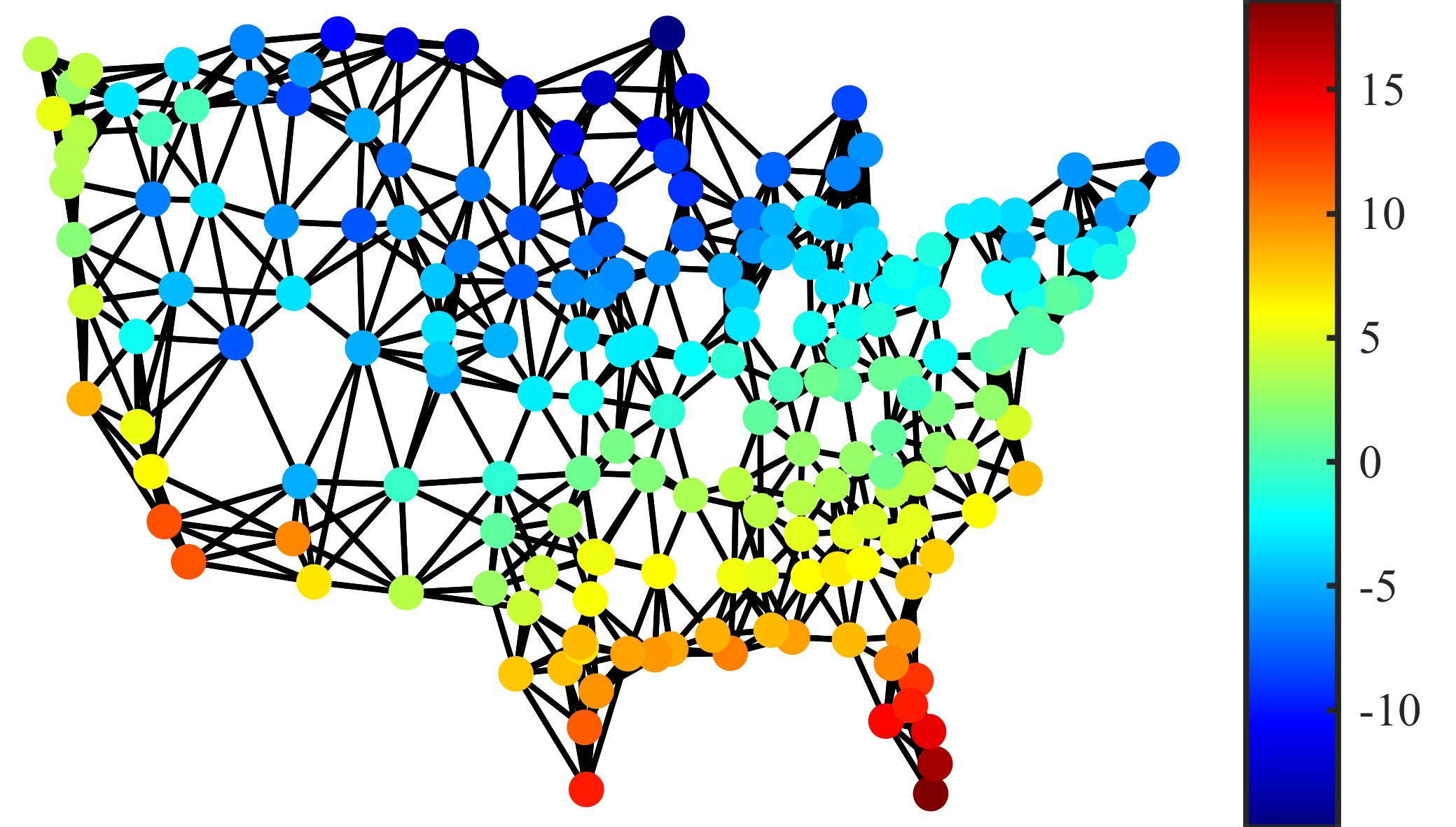}}
\caption{The first time instance of a real time-varying graph signal and its graph topology.}
\label{top_2}
\end{figure}

We would like to test the performance of the G-Sign algorithm on estimating graph signal under impulsive noise. 
Steady-state experiments are conducted in Section~\ref{sec:mu_v} and Section~\ref{sec:steady_state} using the random sensor graph generated by Python package PyGSP shown in Fig.~\ref{top_1} with N = 50, bandlimited frequencies $|\mathcal{F}| = 20$, and greedy sampling strategy in \cite{bib_NLMS} with $|\mathcal{S}|$ = 30. 
A real time-varying graph signal with the topology shown in Fig.~\ref{top_2} is being estimated in ~\ref{sec_tv}. We use geography-based graph generation with 8 nearest neighbors seen in \cite{bib_NLMS} to form the topology shown in Fig.~\ref{top_2} with $N = 205$. 
In the experiment, the sampling technique is the same greedy strategy as in \cite{bib_NLMS} with $|\mathcal{S}|=130$ and $|\mathcal{F}| = 125$. 
The graph signal in Fig.~\ref{top_2} represents hourly temperature recorded across the U.S. at different locations \cite{b22}. 
All the experiments are averaged over 100 independent runs. The experiments were conducted in MATLAB 2020b on a computer with AMD Ryzen 5 3600 CPU and 32GB of RAM.

\subsection{The Effect of the Step-Size Parameter}
\label{sec:mu_v}
\begin{figure}[htbp]
    \centering
    \begin{subfigure}{0.485\textwidth}
        \centering
        \includegraphics[width=\textwidth]{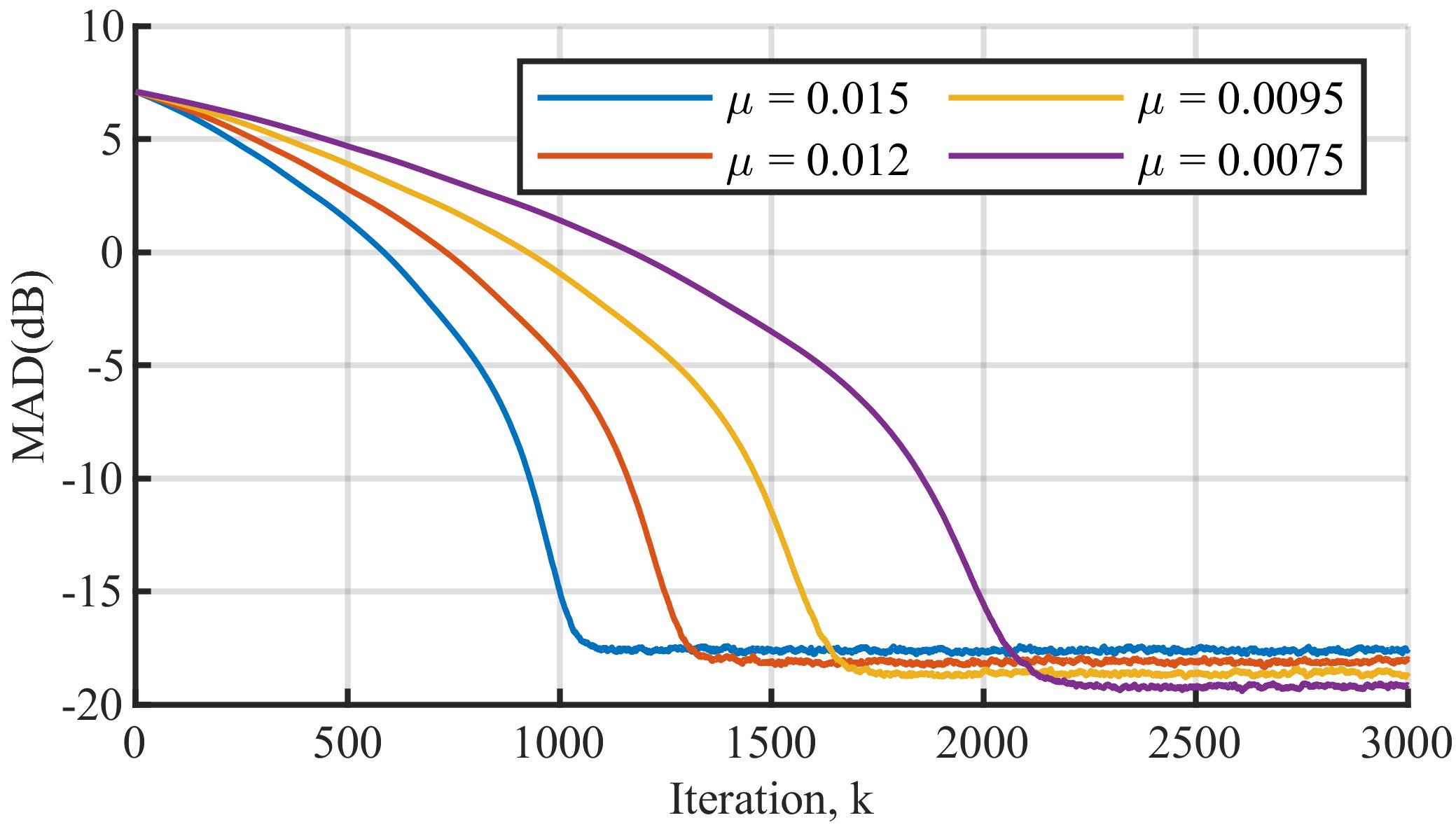}
        \caption{MAD}
        \label{mu_v_MAD}
    \end{subfigure}
    \begin{subfigure}{0.485\textwidth}
        \centering
        \includegraphics[width=\textwidth]{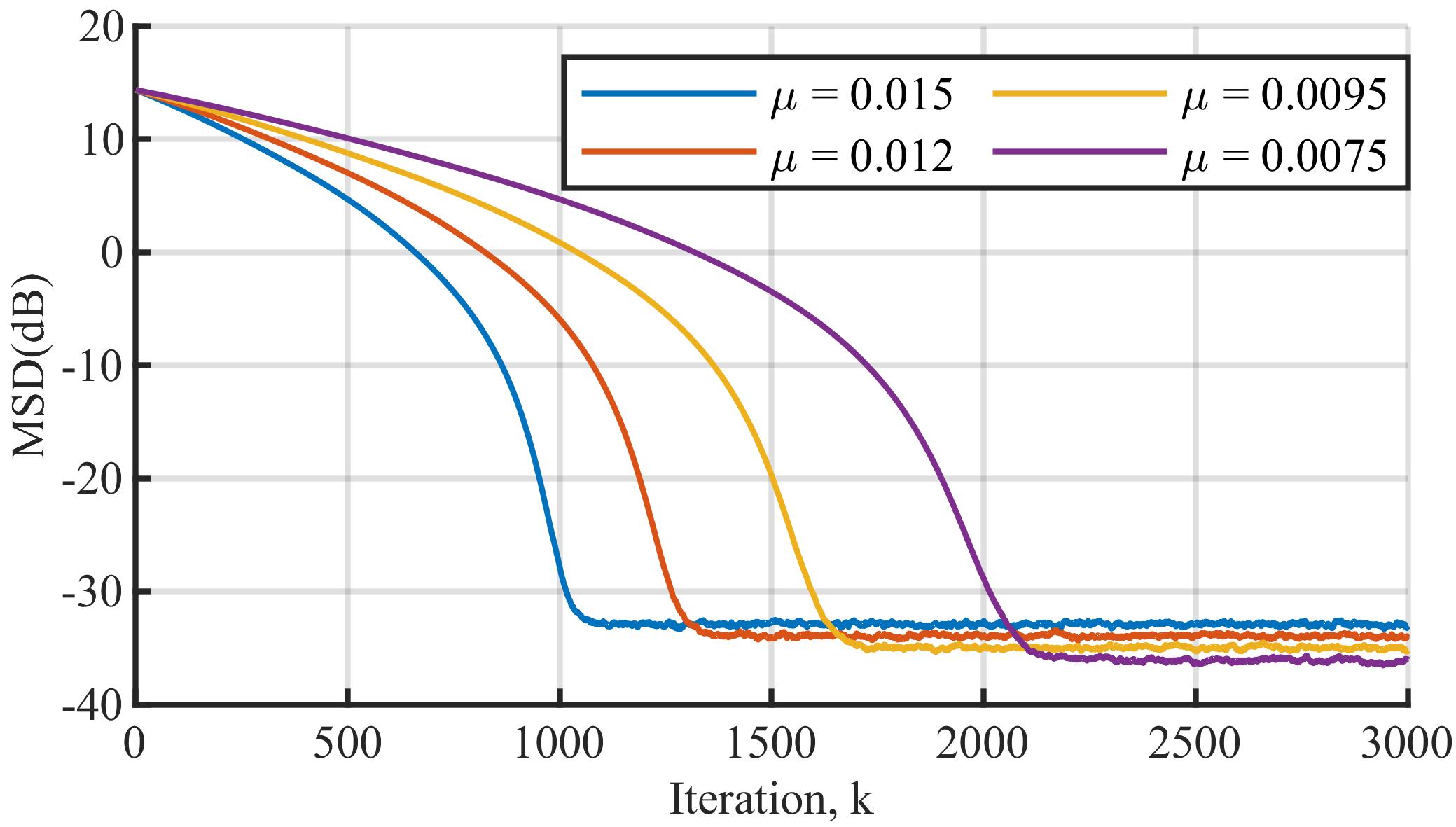}
        \caption{MSD}
        \label{mu_v_MSD}
    \end{subfigure}
    \caption{MAD and MSD performance for estimating a steady-state graph signal under S$\alpha$S noise using different $\mu_s$.}
    \label{mu_v}
\end{figure}

The step-size $\mu_{s}$ is the only user-defined parameter, we would like to see the effect of changing the value of $\mu_{s}$ in the G-Sign algorithm. 
The experiment is being conducted using the graph signal shown in Fig.~\ref{top_1}, and the graph signal is being corrupted by S$\alpha$S noise with $\alpha=1.06$ and $\gamma = 0.1$. 
We tested four different values of $\mu_{s}$; the MAD and MSD performances are shown in Fig.~\ref{mu_v}. We can see that as $\mu_{s}$ decrease, the G-Sign algorithm will get more accurate but will also require more iterations to converge to a steady value.
\subsection{Steady-State Graph Signal Estimation Under Non-Gaussian Noise }
\label{sec:steady_state}
\begin{figure}[!htb]
    \centering
    \begin{subfigure}{0.485\textwidth}
        \centering
        \includegraphics[width=\textwidth]{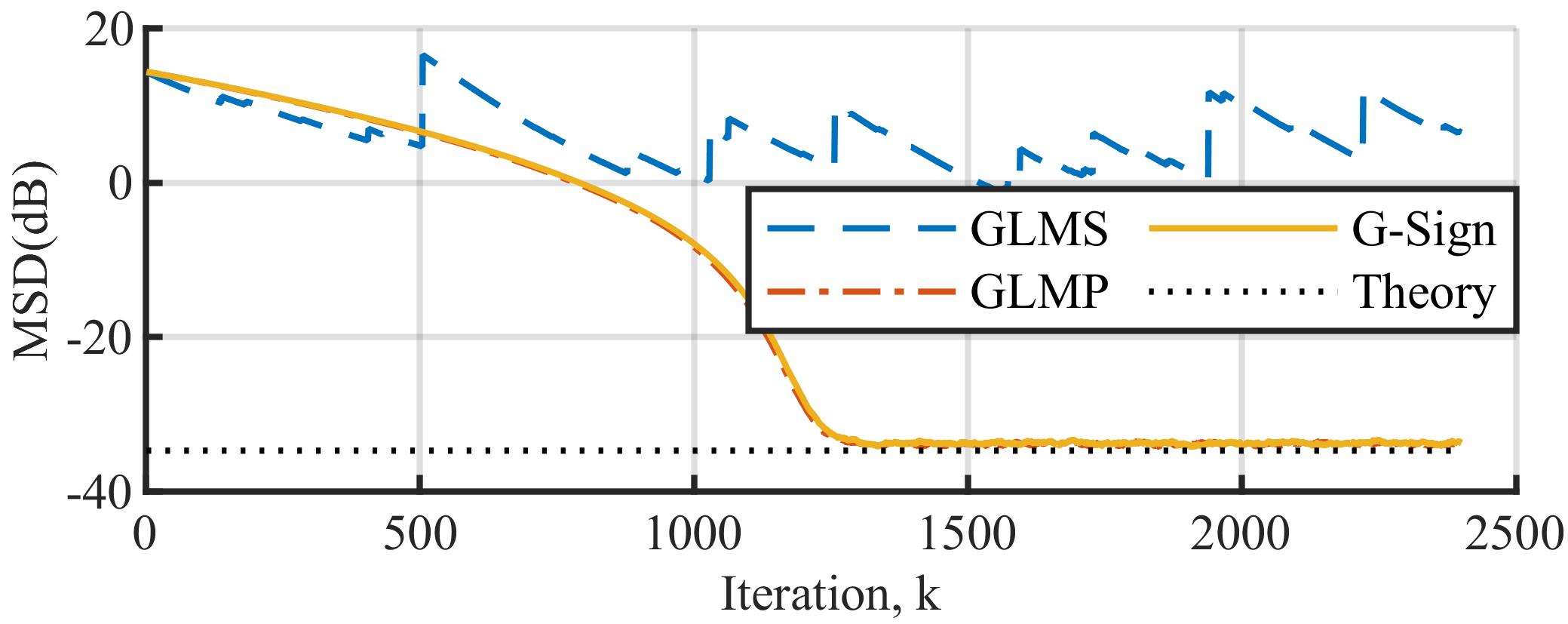}
        \caption{S$\alpha$S noise with $\alpha=1.06$(Near Cauchy) and $\gamma = 0.1$}
        \label{near_cauchy_MSD}
    \end{subfigure}
    \begin{subfigure}{0.485\textwidth}
        \centering
        \includegraphics[width=\textwidth]{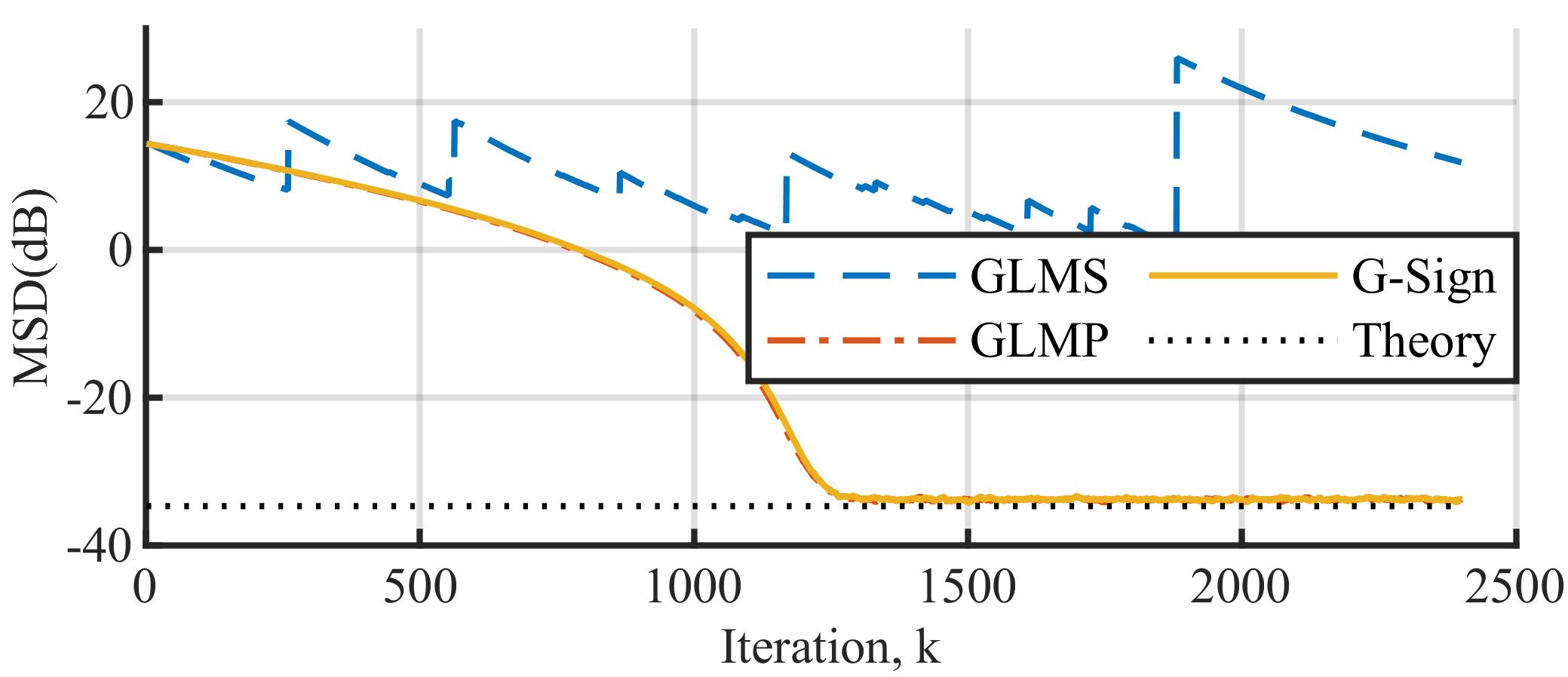}
        \caption{Cauchy with $\mu = 0$ and $\gamma = 0.1$}
        \label{cauchy_MSD}
    \end{subfigure}
    \vskip 0\baselineskip
    \begin{subfigure}{0.485\textwidth}
        \centering
        \includegraphics[width=\textwidth]{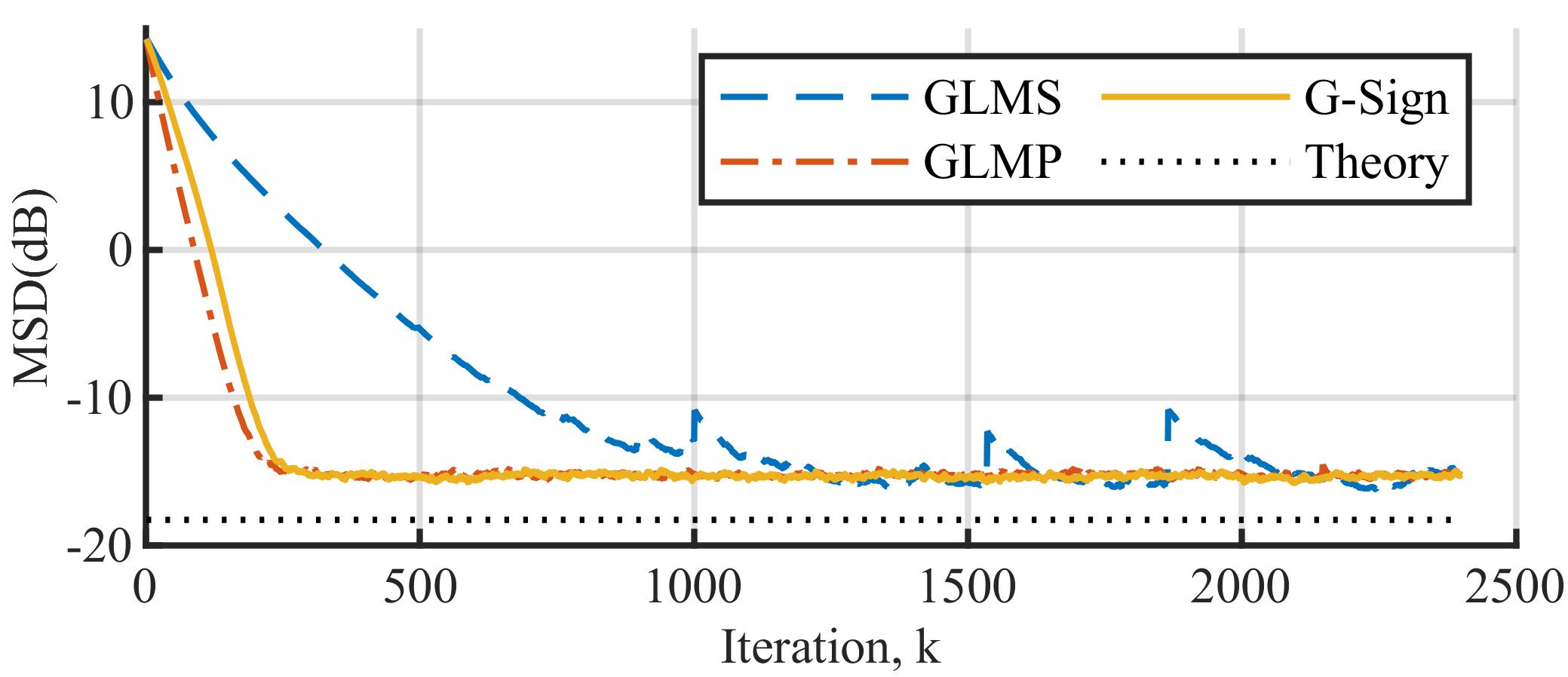}
        \caption{Student's t with $\nu = 2$}
        \label{student_t_MSD}
    \end{subfigure}
    \begin{subfigure}{0.485\textwidth}
        \centering
        \includegraphics[width=\textwidth]{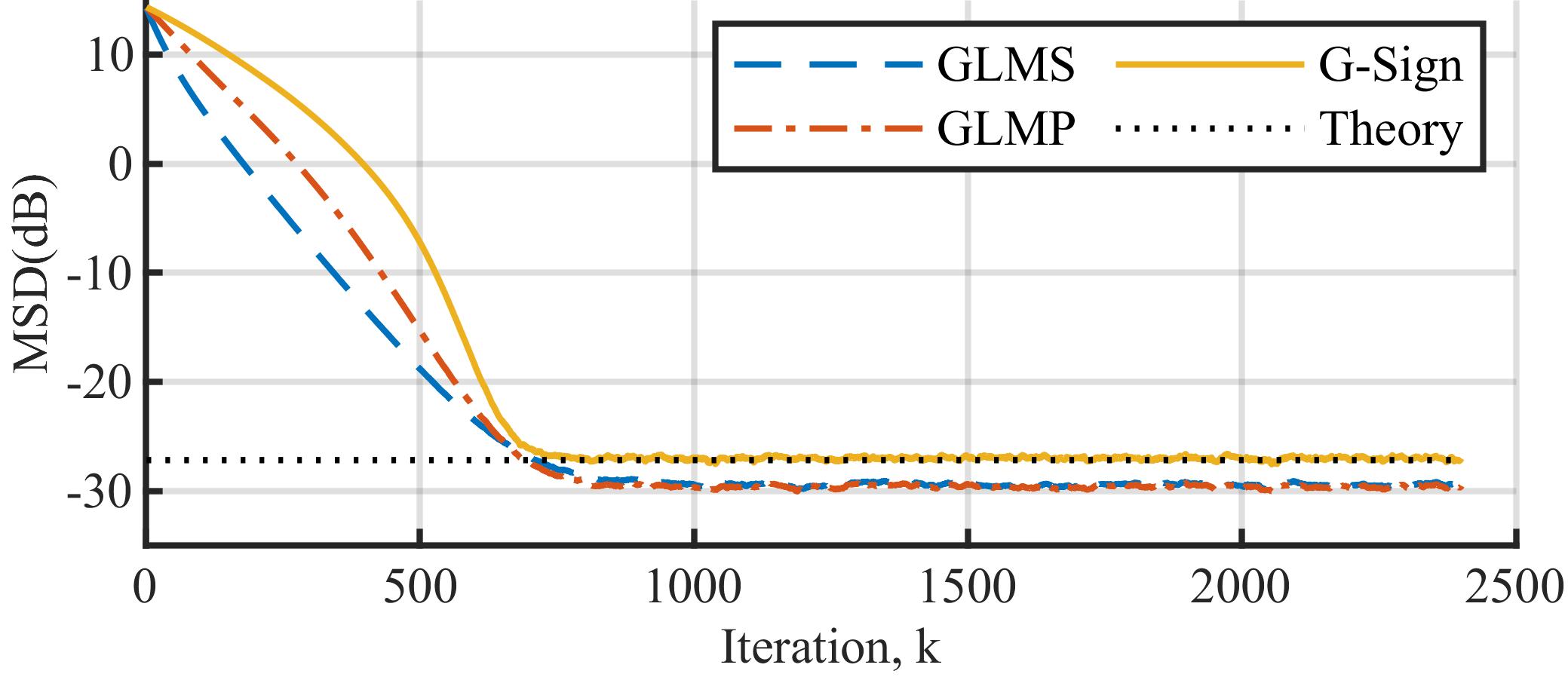}
        \caption{Laplace with $\mu = 0$ and $b = \sqrt{2}$}
        \label{laplace_MSD}
    \end{subfigure}
    \caption{MSD performance for estimating a steady-state graph signal under different noises and the theoretical MSD.}
    \label{steady}
\end{figure}

The G-Sign algorithm is being compared with the GLMS and GLMP algorithms for estimating a partially observed steady-state graph signal under near Cauchy S$\alpha$S, Cauchy, Student's t, and Laplace noises.
The aim is to compare the stability of estimation, the iterations until convergence, and the run-time. In order to fairly compare the algorithms under each noise scenario, the step-sizes are tuned so the algorithms behave similarly in MSD when there is a stable estimation.  
Notice that S$\alpha$S becomes Cauchy when $\alpha$ = 1, and the $p$ parameter of the GLMP algorithm is defined only for $1<p<2$ with $p =\alpha-0.05$ in \cite{bib_LMP}, so we do not test the GLMP algorithm under Cauchy noise. 
Instead, we perform an experiment of near Cauchy S$\alpha$S noise where we set $\alpha = 1.06$ and $p$ = 1.01 for the GLMP algorithm. 
For Laplace noise and Student's t noise, we set $p$ = 1.5 for GLMP. 
The MSD of the experiments is in Fig.~\ref{steady} with the theoretical MSD using \eqref{MSD_simulation}. The run-time of running 2400 iterations of each algorithm for different experiments is in Table \ref{table_time}. 
From Fig.~\ref{steady}, we can see that the GLMS algorithm is unstable when estimating the graph signal under S$\alpha$S, Cauchy, and Student's t noise. 
This instability is introduced by the heavy tail behavior of the noises \cite{bib_nonGaussian}. 
In Fig.~\ref{laplace_MSD}, the GLMS algorithm is stable but requires about 2 times the number of iterations that the G-Sign algorithm needs to converge. 
The G-Sign algorithm behaves similarly to the GLMP algorithm because both algorithms are derived based on the minimum dispersion criterion. 
The MSD performance of the G-Sign algorithm matches the theoretical results under Cauchy, S$\alpha$S, and Laplace noises. 
The gap between theoretical and the actual under Student's t noise is due to slightly lower theoretical MSD caused by the approximation in \eqref{sign_error_2}.
In Table \ref{table_time}, we see that the G-Sign algorithm has the fastest run-time under all scenarios, which is in correspondence with the analysis in Table \ref{complexity} that G-Sign algorithm has the lowest computational complexity. Combining with Fig.~\ref{steady}, we conclude that for steady-state graph signal estimation with missing node values under non-Gaussian noise, the proposed G-Sign algorithm is able to make a stable estimation and faster run-time compared to the GLMS algorithm and the GLMP algorithm. 
\begin{table}[htbp]
    \caption{Run Time Comparison of Steady-State Experiments}
    \label{table_time}
    \centering
    \begin{tabular}{ccccc}
    \hline
         & Near Cauchy & Cauchy & Student's t & Laplace\\
    \hline
        GLMS & 0.0315(s) & 0.0313(s) & 0.0275(s) & 0.0292(s)\\
        GLMP & 0.0476(s) & - & 0.0345(s) & 0.0339(s)\\
        G-Sign &  \bf{0.0061}(s) & \bf{0.0062(s)} & \bf{ 0.0055(s)} & \bf{0.0058(s)}\\
    \hline
    \end{tabular}
\end{table}

\subsection{Time-varying Estimation Under Impulsive Noise}
\label{sec_tv}
\begin{figure}[!htb]
    \centering
    \begin{subfigure}{0.485\textwidth}
        \centering
        \includegraphics[width=\textwidth]{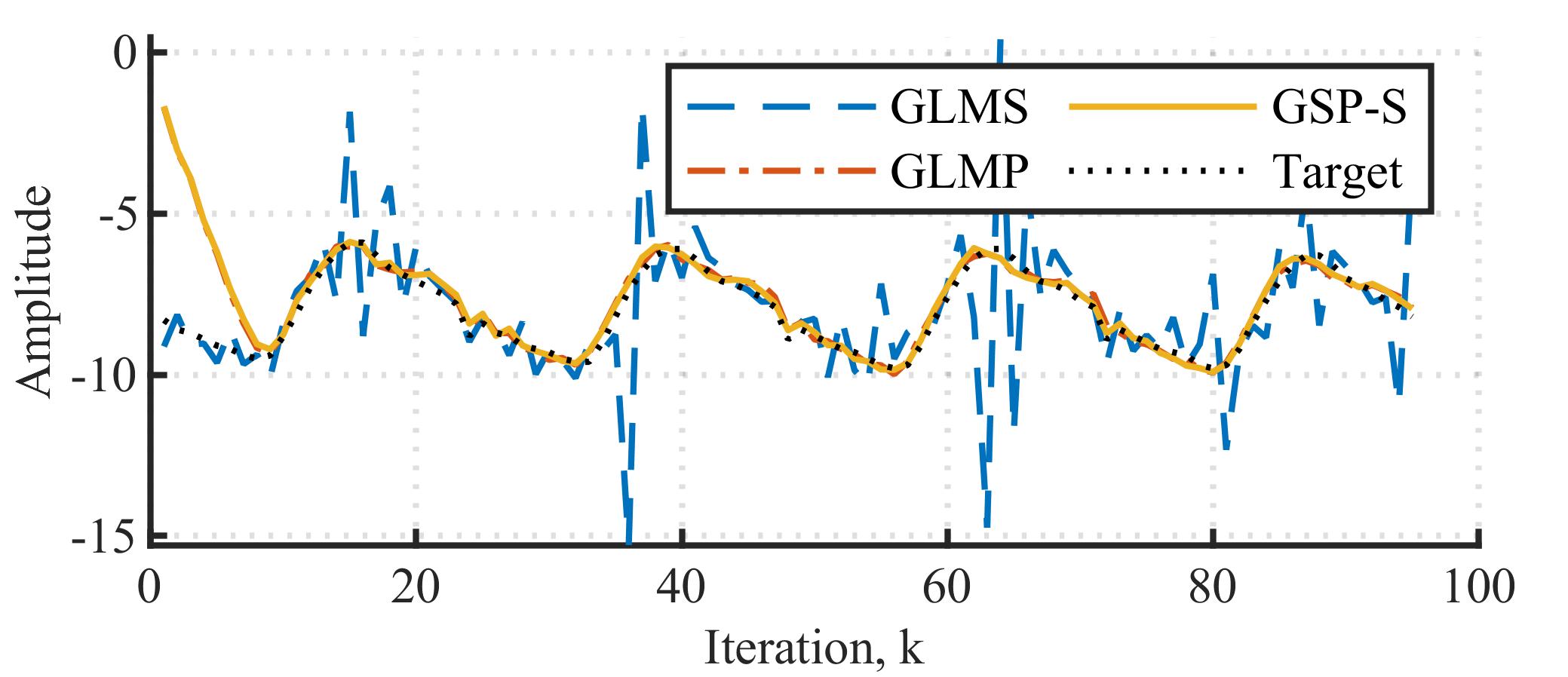}
        \caption{S$\alpha$S noise with $\alpha=1.06$(Near Cauchy) and $\gamma = 0.1$}
        \label{near_cauchy_tv}
    \end{subfigure}
    \begin{subfigure}{0.485\textwidth}
        \centering
        \includegraphics[width=\textwidth]{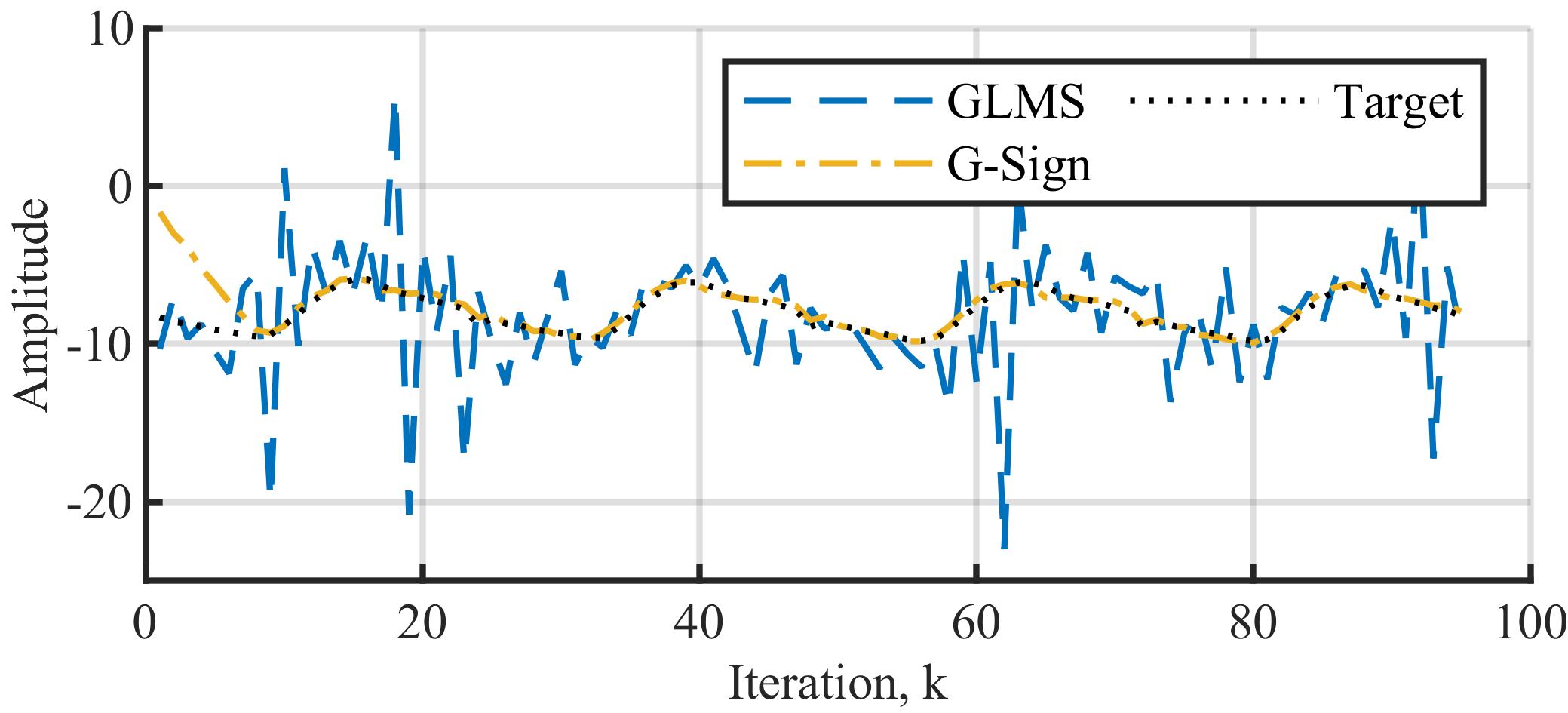}
        \caption{Cauchy with $\mu = 0$ and $\gamma = 0.1$}
        \label{cauchy_tv}
    \end{subfigure}
    \vskip 0\baselineskip
    \begin{subfigure}{0.485\textwidth}
        \centering
        \includegraphics[width=\textwidth]{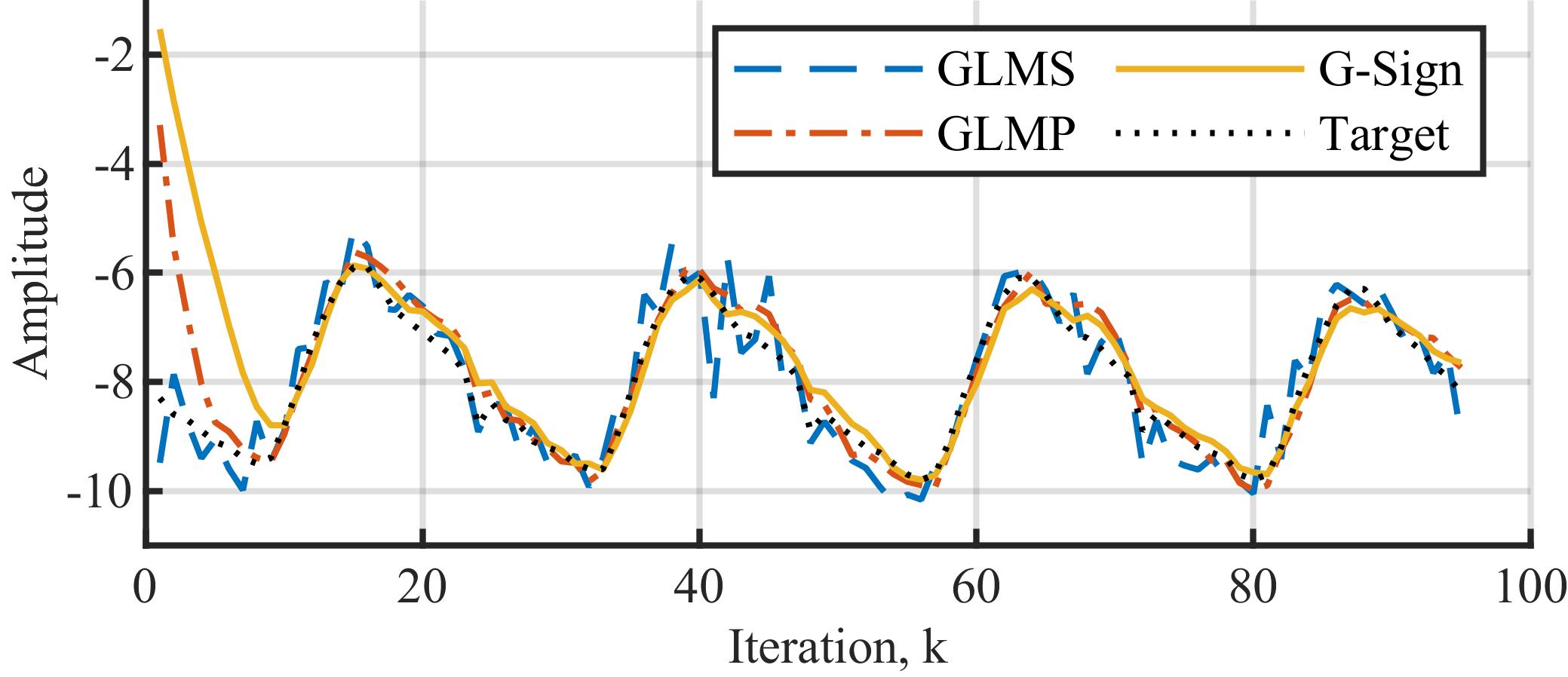}
        \caption{Student's t with $\nu = 2$}
        \label{student_t_tv}
    \end{subfigure}
    \begin{subfigure}{0.485\textwidth}
        \centering
        \includegraphics[width=\textwidth]{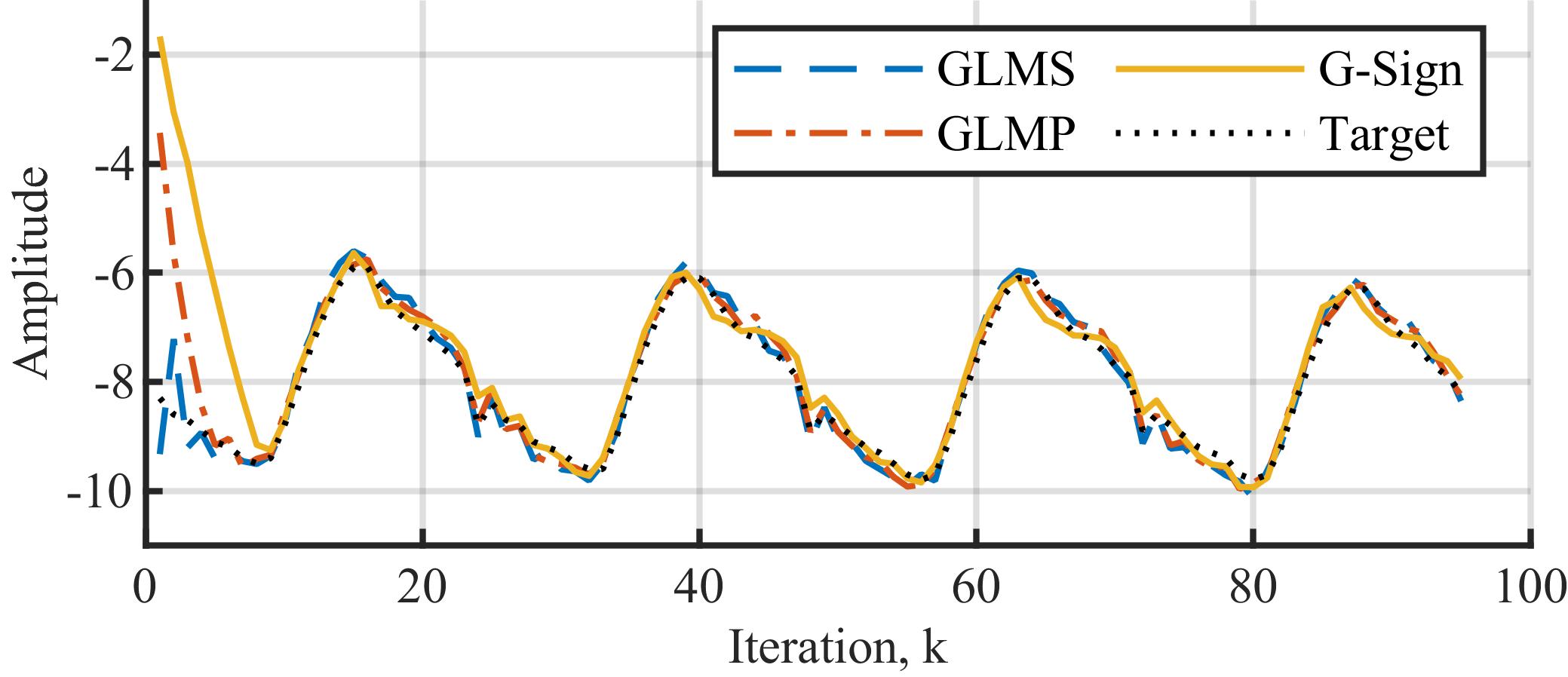}
        \caption{Laplace with $\mu = 0$ and $b = \sqrt{2}$}
        \label{laplace_tv}
    \end{subfigure}
    \caption{Estimation one selected node from a time-varying graph signal under different noises.}
    \label{tv}
\end{figure}

In this section, the G-Sign algorithm will be estimating a time-varying graph signal corrupted by noises modeled by S$\alpha$S, Cauchy, Student's t, and Laplace distributions. The G-Sign algorithm is being compared to the GLMP and GLMS algorithms. 
An illustration of one time step of the graph signal is shown in Fig.~\ref{top_2}. To make a fair comparison, the step-sizes are $\mu_s = 1.5$ for all the algorithms.
Fig.~\ref{tv} illustrates the estimation of one selected node with a time-varying graph signal. 
Notice that the GLMS algorithm is again unstable under S$\alpha$S, Cauchy, and Student's t noises, whereas the G-Sign algorithm is not influenced by any of these impulsive noises. The run-times for GLMS, GLMP, and G-Sign algorithms to finish this experiment are shown in Table \ref{table_time_tv}. 
From Table \ref{table_time_tv}, we can see that the G-Sign algorithm remains the fastest among all compared algorithms under the time-varying setting. 
From Fig.~\ref{tv} and the run-time comparisons, we conclude that the G-Sign algorithm is able to track a time-varying graph signal under non-Gaussian noise in a time-efficient manner.
\begin{table}[htbp]
    \caption{Run Time Comparison of Time-Varing Experiments}
    \centering
    \begin{tabular}{ccccc}
    \hline
         & Near Cauchy & Cauchy & Student's t & Laplace\\
    \hline
        GLMS & 0.0173(s) & 0.0166(s) & 0.0169(s) & 0.0170(s)\\
        GLMP & 0.0194(s) & - & 0.0192(s) & 0.0189(s)\\
        G-Sign &  \bf{0.0029}(s) & \bf{0.0028(s)} & \bf{ 0.0028(s)} & \bf{0.0027(s)}\\
    \hline
    \end{tabular}
    \label{table_time_tv}
\end{table}
\section{Conclusion}
In this paper, we proposed the G-Sign algorithm for online estimation of partially observed steady-state and time-varying graph signals under impulsive noise. 
The G-Sign algorithm is derived using the minimum dispersion criterion which is stable and robust under impulsive noise. 
Experimental results confirm that the G-Sign algorithm is low complexity, time-efficient and robust.
\bibliography{references}

\end{document}